\documentclass[aps,reprint,superscriptaddress,nofootinbib,showkeys,longbibliography]{revtex4-2}

\usepackage{graphicx}
\usepackage{amsmath,amssymb}
\usepackage[colorlinks,linkcolor=blue,citecolor=blue]{hyperref}

\newcommand \Title{
    Log-normal Superstatistics Reveals Statistical Resilience\\in the Panic Response of Confined Ants
}

\newcommand{\affilifshown}[1]{\ifdefined\noaffil
\else
    \affiliation{#1}\fi
}

\makeatletter
\let\oldsection\section

\newcommand{\section@nostar}[2][]{\oldsection[#1]{#2}\vspace*{-0.8\baselineskip}}

\newcommand{\section@star}[1]{\oldsection*{#1}\vspace*{-0.8\baselineskip}}

\renewcommand{\section}{\@ifstar{\section@star}{\@dblarg\section@nostar}}
\makeatother

\newcommand \WallExtent{0.6}
\newcommand \WallIntensity{1.0}
\newcommand \InvGamma{0.53}
\newcommand \Ssigma{0.34}
\newcommand \DeltaSigma{1.99}
\newcommand \Vc{1.99}
\newcommand \PanicPeakTime{9.2}
\newcommand \DTimeScale{3.3}
\newcommand \RatioTimeScales{6}

\newcommand \ShapeParam{1.27}
\newcommand \CMTimeSlow{20}
\newcommand \CMTimeFast{3.2}
\newcommand \CellLength{6.0}
\newcommand \TStartSigmoid{5.5}
\newcommand \CMSaturationLength{3.8}
\newcommand \PanicDeltaSigmaBase{1.53}
\newcommand \PanicDeltaSigmaAmp{3.07}
\newcommand \PanicDeltaSigmaRate{0.09}
\newcommand \PanicVcBase{1.25}
\newcommand \PanicVcAmp{5.48}
\newcommand \PanicVcRate{0.06}

\newcommand \Units[1]{\,\mathrm{#1}}
\newcommand{\dexp}[1]{e^{\displaystyle{#1}}}

\begin{document}

\title{\Title}

\author{A. Reyes}
\email{areyes@fisica.uh.cu}
\affilifshown{Center for Complex Systems, Physics Faculty, University of Havana, 10400 Habana, Cuba}

\author{M. Curbelo}
\affilifshown{Biology Faculty, University of Havana, 10400 Havana, Cuba}

\author{F. Tejera}
\affilifshown{The Rockefeller University, New York, USA}

\author{A. Rivera}
\affilifshown{Zeolites Engineering Lab, University of Havana, 10400 Havana, Cuba}

\author{M.S. Turner}
\affilifshown{Physics Department \& Centre for Complexity Science, University of Warwick, Coventry, CV4 7AL, UK}

\author{O. Ramos}
\affilifshown{Institut Lumi\`ere Mati\`ere, UMR5306, Universit\'e Claude Bernard Lyon 1, France}

\author{E. Altshuler}
\email{ealtshuler@fisica.uh.cu}
\affilifshown{Center for Complex Systems, Physics Faculty, University of Havana, 10400 Habana, Cuba}
 
\begin{abstract}

We report the emergence of Log-normal Superstatistics in the collective motion of ants confined in a quasi-2D arena and exposed to a panic-inducing stimulus. A data-driven superstatistical Langevin model accurately reproduces the transition from stationary behavior to an organized escape response, characterized by non-Gaussian velocity distributions and a stochastic diffusion coefficient. Our findings show that danger information propagates via a memory-limited, cascade-like mechanism, resulting in a stable cluster formation despite individual memory constraints. These results indicate that a slowly varying diffusivity arises from the multiplicative combination of interaction-mediated processes under confinement, leading naturally to Log-normal fluctuations. The persistence of this statistical structure under panic reveals a form of collective resilience, establishing a mechanistic bridge between Superstatistics and living active matter in confined environments.

\end{abstract}
 
\maketitle

\section{Introduction}

Complex systems often exhibit emergent behavior that challenges classical Statistical Mechanics. In particular, far-from-equilibrium systems can feature fast local dynamics that coexist with slow fluctuations of intensive parameters such as temperature or diffusivity~\cite{beck2001dynamical,beck2003superstatistics,wang2009anomalous,metzler2020superstatistics,chechkin2017brownian}. Superstatistics (i.e., a superposition of different local equilibrium statistics) has been successfully applied to describe these frameworks with strong parameter fluctuations, such as turbulence, convection and active particles~\cite{beck2007statistics,queiros2008superstatistical,rizzo2004environmental,Romanczuk2011}. In living matter, devoted studies include DNA architecture and the motion of proteins and tumor cells~\cite{bogachev2017superstatistical, Itto2021, Metzner2015}. Although superstatistical ideas have been theoretically connected to macroscopic biological systems~\cite{santana2020langevin}, their dedicated experimental application to animal movement is still developing~\cite{costa2024fluctuating}. The collective behavior of animals, particularly eusocial insects like ants, presents an ideal substrate for applying superstatistical frameworks. Colonies of ants respond to environmental stimuli using distributed sensing and local communication, displaying pattern formation, foraging trails, and escape dynamics~\cite{holldobler1990ants,gordon2010ant,sumpter2006principles,bonabeau1997self}.

A canonical starting point for modeling directional animal movement is the Langevin equation~\cite{viswanathan2011physics,mendez2016stochastic,gurarie2017correlated,hadeler2004langevin}, which describes the $i$-th Cartesian velocity component $v_i$ of an individual as a balance between a frictional term and a stochastic noise. In its simplest form, it reads:
\begin{equation}
	\frac{d}{dt} v_i(t) = -\gamma v_i(t) + \sigma \xi_i(t)
	\label{eq:langevin-model}
\end{equation}

\noindent
where $\gamma$ is a positive parameter that encloses directional persistence, and $\xi_i$ is a Gaussian white noise of intensity $\sigma$. Superstatistics extends this view by allowing intensive parameters such as $\beta = 2\gamma / \sigma^2$ to fluctuate on longer time scales. The system is assumed to reach local equilibrium over a short time $\tau_1$, while the parameter $\beta$ varies over a slower scale $\tau_2 \gg \tau_1$. Consequently, within an interval of duration $\tau_2$, $\beta$ can be treated as approximately constant and the conditioned velocity distribution takes the Gaussian form:
\begin{equation}
	p(v_i|\beta) = \left(\frac{\beta}{2\pi}\right)^{1/2}
	\dexp{-\beta v_i^2 / 2}
\end{equation}	

Therefore, to capture the long-time behavior, one integrates over the ensemble of local equilibria weighted by the distribution $f(\beta)$, yielding the marginal density for the $i$-th Cartesian velocity component:
\begin{equation}
	f(v_i) = \int_0^\infty p(v_i|\beta) f(\beta) \,d\beta
\end{equation}
	
\noindent
which yields diverse non-Gaussian stationary distributions documented in many physical systems~\cite{porporato2006superstatistics,briggs2007modelling,chen2008superstatistical,abul2006superstatistics,abe2005complex,dos2020log}.

Here, we ask whether a fundamental statistical framework underlies ants movement in confinement and persists or adapts when ants are perturbed by panic stimuli. We show that such a framework aligns to the principles of Superstatistics and study the motion of confined ants as a case study for superstatistical analysis. To this end, we first describe the experimental setup and the emergence of the collective escape response under confinement. We then analyze a baseline regime to construct a superstatistical description of single-ant motion, and next examine how this framework extends to the panic regime and the resulting cluster dynamics. Finally, we discuss the persistence of the underlying statistical structure under perturbation and its implications for collective resilience in living active matter. The proposed superstatistical model captures both regimes, providing an experimental report of Superstatistics in the animal kingdom.
 \section{Experimental details}

Experiments were performed in dual acrylic arenas, labelled danger and safe, of $6\times6\times1\Units{cm^3}$ connected via a \textit{Thin} barrier designed to allow contacts of different types (see Appendix B) (Fig.~\ref{fig:exp-setup}a). Each cell was populated with nearly 20 adult workers of the Cuban ant species \textit{Atta insularis} (Fig.~\ref{fig:atta-insularis}a), randomly selected from a single colony and previously studied in both natural and laboratory settings~\cite{nicolis2013foraging,noda2006measuring,tejera2016uninformed,serrano2019autonomous}. This density is consistent with crowded nest regions and dense foraging lines~\cite{bruce2012allometric,dussutour2009priority}, and matches occupancy regimes in which collective escape asymmetries and panic-induced effects have previously been reported under confinement~\cite{altshuler2005symmetry,li2014symmetry}. After a 5-minute baseline, $50 \Units{\mu L}$ of insect repellent was introduced into the danger cell. Video data were recorded for a total of 15 minutes at 30 fps. In control (\textit{Ctrl}) experiments no ants were introduced into the danger arena. Each protocol was repeated 10 times using different individuals to avoid memory effects~\cite{gruter2019communication}.

A significant shift of the center of mass (CM), measured via simple video binarization, away from the barrier was observed after introducing the repellent, indicating an escape response (Fig.~\ref{fig:exp-setup}b,d). The negligible shift in the \textit{Ctrl} protocol (Fig.~\ref{fig:exp-setup}c,e) confirmed that this effect was not due to passive diffusion of the repellent and is consistent with interaction-mediated information transfer across the two arenas. Fig.~\ref{fig:exp-setup}d shows that the CM velocity increases sharply after repellent exposure, indicating a rapid initial reaction followed by a slower phase in which ants form a stable cluster away from the barrier. This two-stage process, i.e., early activation followed by clustering, will be discussed later. Additional protocols, time-resolved dynamics, statistical validation, and experiment videos with CM evolution are provided in Appendixes and Supplementary Information (SI), where we also indicate that optical signals or simple antennation are not significant mechanisms for panic transmission (Fig.~\ref{fig:cm-displacements}).

We now show that single-ant velocities and accelerations can be analyzed and incorporated into a Langevin-based model that accurately reproduces the ensemble's CM dynamics in the safe cell before repellent injection in \textit{Ctrl} experiments, where ants interact only with each other and the barrier, free from influence by ants in the danger arena (baseline regime). This unperturbed state establishes a reference model of ant behavior under confinement based on the Langevin equation~\eqref{eq:langevin-model}, which we later extend to incorporate interactions with ants in the danger cell and the transmission of danger information in \textit{Thin} experiments (panic regime) (Fig.~\ref{fig:exp-setup}d).

\begin{figure}
	\includegraphics[width=\linewidth]{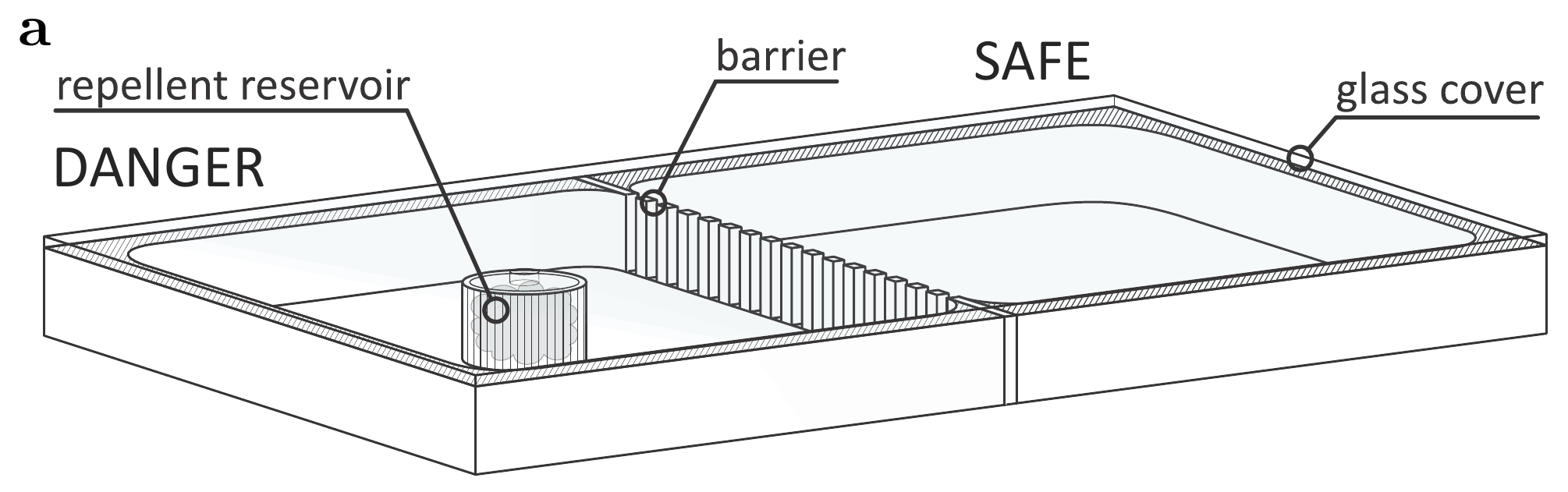}
	\includegraphics[width=\linewidth]{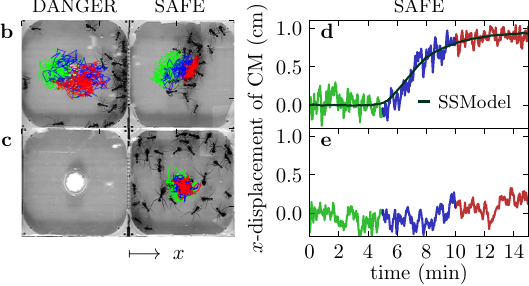}
	\caption{
		Experimental Setup and Collective Escape Response.
		(a) Experimental setup and snapshots of experiments for 
		(b) \textit{Thin} and 
		(c) \textit{Ctrl} protocols with the center of mass (CM) trajectory colored by 5-minute intervals (green, blue, red).
		(d) Net horizontal CM displacement relative to its initial position for the \textit{Thin} protocol, as predicted by the superstatistical model in Eq.~\eqref{eq:panic-langevin-model} (SSModel);
		(e) corresponding experimental data for the \textit{Ctrl} protocol.}
	\label{fig:exp-setup}
\end{figure}
 \section{Baseline regime} 

Baseline regime was characterized by tracking 18 randomly selected ants, using the yupi library for trajectory analysis~\cite{reyes2023yupi}. The sample size was chosen to ensure reliable averages, targeting $\sim 20$ individuals as a compromise between minimal small-sample statistics ($\sim 10$) and the regime where ensemble averages begin to approach Gaussian convergence ($\sim 30$), and trajectories were selected by uniform random sampling among ants visible in the safe cell at the beginning of the baseline window across the 10 repetitions. To avoid boundary effects, only trajectory segments beyond $\ell_w = \WallExtent \Units{cm}$ from all walls were used in the analysis, where $\ell_w$ is a characteristic length obtained from the exponential decay of ant acceleration near walls (Fig.~\ref{fig:boundary-effect}).

\begin{figure*}
	\includegraphics[width=\linewidth]{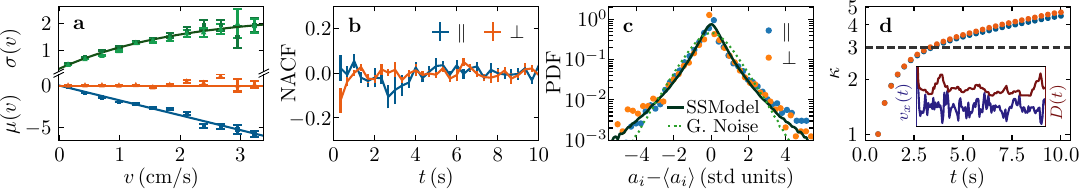}
	\caption{
		Speed-Dependent Fluctuations and Time-Scale Separation in the Baseline Regime.
		(a) Mean (blue/orange) and fluctuation intensity (dark/light green) of parallel and perpendicular acceleration components vs. speed. Error bars account for the standard error of the mean. Parallel acceleration scales linearly, perpendicular remains near zero, and both speed-dependent standard deviations fit Eq.~\eqref{eq:sigma-in-v}. Note that $\mu(v)$ and $\sigma(v)$ differ in units (i.e., $\mathrm{cm \, s}^{-2}$ and $\mathrm{cm \, s}^{-3/2}$, respectively).
(b) Autocorrelation function of acceleration fluctuations around their mean (Noise Autocorrelation Function, NACF). Parallel and perpendicular components are shown in blue and orange, respectively. The fluctuations exhibit no temporal correlation, indicating uncorrelated noise.
(c) Probability density function (PDF) of parallel and perpendicular acceleration fluctuations. The non-Gaussian shape is accurately captured by the SSModel (Eq.~\ref{eq:ss-model}), which combines a Log-normal and Gaussian dynamics, but it cannot be explained only by Gaussian noise (G. Noise) with the state-dependent intensity $\sigma(v)$.
(d) Time-average of local flatness of Cartesian velocity components (Eq.~\eqref{eq:flatness}). The dashed line marks the Gaussian value. Inset: typical realization of $D(t)$ (dark red) varies more slowly than $v_x(t)$ (dark blue) over $\sim 50 \Units{s}$.}
	\label{fig:baseline-regime1}
\end{figure*}

To gain theoretical insight into the persistence of ant motion and both the intensity and the nature of fluctuations,
we projected acceleration onto a velocity-based moving frame defined by unit vectors parallel and perpendicular to the direction of motion: $\mathbf{e}^{(\|)} = \mathbf{v}/v$ and $\mathbf{e}^{(\bot)} = (-e^{(\|)}_y, e^{(\|)}_x)^\top$. We then computed, as functions of speed, the mean parallel acceleration $\mu(v) \equiv \langle a^{(\|)} \rangle$ and the fluctuation intensities $\sigma(v) \equiv \langle (a^{(\|,\bot)} - \langle a^{(\|,\bot)} \rangle)^2 \rangle^{1/2} \Delta t^{1/2}$ for both components. The latter was scaled by $\Delta t^{1/2}$ to ensure dimensional consistency with a continuous stochastic process, where the white noise has units of $\mathrm{s}^{-1/2}$~\cite{gardiner1985handbook}.
Throughout the manuscript, we reserve ``parallel'' and ``perpendicular'' for this instantaneous moving frame, while $x$ and $y$ refer exclusively to the fixed Cartesian frame of the arena.

Fig.~\ref{fig:baseline-regime1}a shows that the acceleration component parallel to the velocity averages to $\mu(v) = -\gamma v$, where $\gamma^{-1} = \InvGamma \Units{s}$. The perpendicular component averages to zero across speeds. However, the intensity of fluctuations in both components depends nonlinearly on speed and follows the same trend within their error bars. We model these behavior with
\begin{equation}
	\sigma(v) = \Sigma + \Delta\Sigma \, (1 - \dexp{-v / v_\sigma})
	\label{eq:sigma-in-v}
\end{equation}

\noindent
where $\Sigma$ captures resting fluctuation intensity, $\Delta\Sigma$ sets the total increase in noise intensity from rest to high speeds, and $v_\sigma = \Vc\Units{cm/s}$ modulates the speed dependence. Although alternative forms (e.g., power laws) were tested, the exponential fit performed best, avoiding unphysical values like $\sigma(0) \le 0$. Furthermore, the temporal correlation of ant velocities decays exponentially (Fig.~\ref{fig:after-panic-results}b), yielding a persistence time $\gamma^{-1}$ consistent with the slope of the parallel acceleration.

The acceleration fluctuations show negligible temporal correlations (Fig.~\ref{fig:baseline-regime1}b). However, Fig.~\ref{fig:baseline-regime1}c shows that they deviate from the Gaussian noise assumed in Eq.~\eqref{eq:langevin-model}. Although state-dependent noise $\sigma(v)\,\xi_i(t)$ introduces non-Gaussian features, it does not fully account for the observed tails. A natural extension beyond time-invariant noise intensity is therefore to treat it as a slowly varying stochastic process, placing the system within the realm of Superstatistics. In this picture, the observed non-Gaussian fluctuations arise from a superposition of local Gaussian processes whose variances fluctuate over longer timescales.

To test whether such a separation of timescales exists in the data, we computed the time-dependent flatness $\kappa_i(t)$ of the Cartesian velocity components~\cite{beck2005time}:
\begin{equation}
	\kappa_i(t) =
	\frac{1}{T - t} \int_0^{T - t}
	\frac{
		\langle(v_i - \langle v_i \rangle)^4\rangle_t}{
		\langle(v_i - \langle v_i \rangle)^2\rangle_t^2} \, dt',
	\label{eq:flatness}
\end{equation}
where $T$ is the total measurement time, $\langle \cdot \rangle_t$ denotes averages computed within sliding windows of duration $t$, and $t'$ labels the starting time of the window. The flatness is the normalized fourth moment (kurtosis) and equals 3 for a Gaussian distribution; deviations from this value quantify non-Gaussianity. Here, the window size $t$ plays the role of a coarse-graining timescale: for small $t$, statistics reflect instantaneous fluctuations, while increasing $t$ progressively averages over slower modulations.

The slow time scale $\tau_2$ was extracted as the window size at which the flatness $\kappa_i(t)$ approaches the Gaussian value, indicating that, despite the presence of multiplicative noise, at this coarse-graining scale the local velocity statistics are well approximated by a quasi-Gaussian distribution (Fig.~\ref{fig:baseline-regime1}d). We find $\tau_2 = \DTimeScale \Units{s}$, which is much longer than the velocity decorrelation time $\tau_1 = \gamma^{-1} = \InvGamma \Units{s}$; their ratio $\tau_2 / \tau_1 \approx \RatioTimeScales$ confirms the existence of an intensive parameter $\beta(t)$ that fluctuates on a timescale about \RatioTimeScales{} times slower than the velocity itself.

The slowly varying stochastic process $\beta(t)$ was then computed over sliding windows of length $\tau_2$ as the inverse of the local velocity variance, 
\begin{equation}
	\beta(t) = [\langle v_i^2(t) \rangle_{\tau_2} - \langle v_i(t) \rangle_{\tau_2}^2]^{-1}
\end{equation}

\begin{figure}
	\centering
	\includegraphics[width=\linewidth]{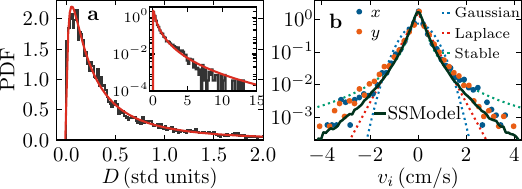}
	\caption{
		Log-normal Diffusivity and Non-Gaussian Velocities in the Baseline Regime.
		(a) Probability density (PDF) of diffusion coefficient $D$ (standardized) fitted to a Log-normal distribution. Inset shows the tail up to 15 standard deviations.
(b) Velocity PDF with Gaussian, Laplace, and power-law (L\'evy stable) fits. The SSModel~\eqref{eq:ss-model} (solid line) captures the central peak and tails beyond Gaussian decay.}
	\label{fig:baseline-regime2}
\end{figure}

\noindent
Consequently, we explored suitable powers $p$ such that $\beta^p$ could make the dynamics fall into one of the universality superstatistical classes (e.g., $\chi^2$ or Log-normal \cite{beck2006stretched}) and found that $p = -1$ yields the diffusion coefficient $D \propto \beta^{-1}$ as the relevant slowly varying quantity. Fig.~\ref{fig:baseline-regime2}a shows that $D$ follows a Log-normal distribution with PDF
\begin{equation}
	f(D) =
	\frac{1}{\sqrt{2\pi s^2}\, D}
	\exp\!\left(
		-\frac{\ln^2 D}{2s^2}
	\right)
	\label{eq:lognormal-d}
\end{equation}
\noindent
where the fitted shape parameter is $s = \ShapeParam$. Although Log-normal and stretched exponential fits can appear similar when tails are limited~\cite{beck2006stretched}, the inset shows a robust Log-normal fit extending over 15 standard deviations. Typical time series of $v_x(t)$ and $D(t)$ in Fig.~\ref{fig:baseline-regime1}d (inset) visually confirm their separate time scales.

Once gathered the measured persistence time $\gamma^{-1}$, the speed-dependent noise intensity $\sigma(v)$, and the slowly varying diffusion coefficient $D(t)$, we set them into a superstatistical Langevin model \cite{beck2003superstatistics} for a Cartesian velocity component $v_i(t)$ in the safe cell before panic onset:
\begin{equation}
    \frac{d}{dt} v_i(t) = 
	- \gamma v_i(t)
	+ \sigma(v) \sqrt{\hat D(t)} \,\xi_i(t)
  \label{eq:ss-model}
\end{equation}

\noindent
where $\hat D(t)$ is introduced through the normalization of the compound noise generated by $D(t)$. More explicitly, what we use from $D(t)$ in the simulations is its Log-normal shape parameter $s$, which determines the statistics of the non-Gaussian noise $\eta(t)=\sqrt{D(t)}\,\xi_i(t)$. This compound noise is then standardized to unit variance, yielding $\hat\eta(t)=\sqrt{\hat D(t)}\,\xi_i(t)$, and finally scaled by $\sigma(v)$. We refer to this as the SSModel.

Numerical integration was performed using a Runge-Kutta method of a strong order 1.5 within a Milstein scheme to compute Stratonovich-type integrals~\cite{burrage1996high}. Fig.~\ref{fig:baseline-regime2}b shows that the SSModel accurately predicts the velocity distribution: no discrepancies are observed in the shapes, consistent with the arena's spatial homogeneity and isotropy. The model reproduces both the low-speed peak and the non-Gaussian tails through the combined effects of $\sigma(v)$ and Log-normal fluctuations in $D(t)$. It also captures the non-Gaussian fluctuations of parallel and perpendicular accelerations (Fig.~\ref{fig:baseline-regime1}c).
 \section{Panic regime} 

Panic regime was examined by tracking 48 trajectories of ants during the next 10 minutes following repellent injection. A larger sample ($\sim 50$ trajectories) was targeted to ensure sufficient statistics for time-resolved estimates, and manual tracking yielded 48 usable trajectories. 

\begin{figure}
	\includegraphics[width=\linewidth]{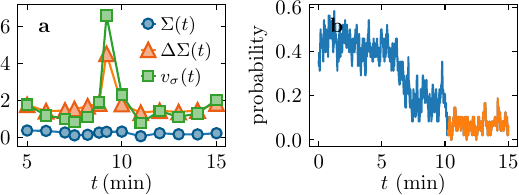}
	\caption{
		Temporal Signatures of Panic and Cluster Stabilization in the Panic Regime.
		(a) Time evolution of the parameters in Eq.~\eqref{eq:sigma-in-v}, obtained from independent fits in successive panic-regime time windows. $\Delta\Sigma(t)$ and $v_\sigma(t)$ show a sharp change at minute \PanicPeakTime, while $\Sigma(t)$ remains stable.
(b) Probability of finding an ant from the safe cell at a distance less than or equal to $2 \Units{cm}$ from the barrier. The last 5 minutes of the experiment are highlighted in orange, where the curve remains approximately constant near zero. During this interval, the ant cluster is stable.}
	\label{fig:panic-regime}
\end{figure}

\begin{figure*}[b!]
\includegraphics[width=\linewidth]{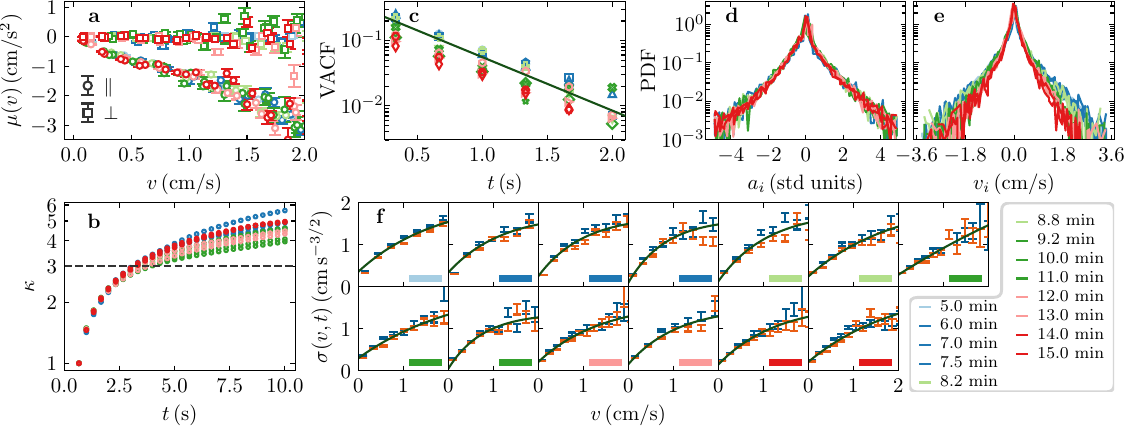}
\caption{
			Persistence of Superstatistical Signatures Under Panic.
			Panels (a-f) were computed over consecutive time segments. The first segment corresponds to the baseline interval $[0,5)$ min, whereas the panic regime was analyzed using shorter windows; the legend reports the final time of each segment, so $[14,15)$ denotes the last $1$-min interval.
(a) Speed dependence of parallel and perpendicular acceleration components across time intervals.
(b) Local flatness of Cartesian velocity components vs lag time, computed over those trajectory segments.
(c) Velocity autocorrelation functions (VACF) vs lag time. The solid line has slope $\gamma = (\InvGamma \Units{s})^{-1}$. Notice that the slow time scale $\tau_2$ is not captured by the VACF.
(d) Probability density function (PDF) of centered parallel and perpendicular components of acceleration fluctuation.
(e) PDF of Cartesian velocity components.
(f) Temporal and speed dependence of the intensity of acceleration fluctuations for the parallel (blue) and perpendicular (orange) components after panic was induced. Continuous lines show fits to Eq.~\eqref{eq:sigma-in-v}.}
	\label{fig:after-panic-results}
\end{figure*}

Several observables remained unchanged (Fig.~\ref{fig:after-panic-results}): mean acceleration stayed linear in speed, $\tau_2$ remained stable, and both $\gamma^{-1}$ and $D$ showed no significant changes. Centered acceleration distributions also retained their shape, indicating preserved local Gaussian equilibria for $t < \tau_2$. However, the velocity distributions developed stretched tails, consistent with gradual clustering in the safe cell, and non-gaussianity persisted due to the interplay of $v$-dependent noise and Superstatistics. The only clearly altered observable was $\sigma(v)$, which maintained isotropy but evolved over time. To quantify this evolution, Eq.~\eqref{eq:sigma-in-v} was fitted independently in successive time windows during the panic regime, yielding one triplet $(\Sigma,\Delta\Sigma,v_\sigma)$ per window. Figure~\ref{fig:panic-regime}a shows the resulting scalar time series: $\Delta\Sigma(t)$ and $v_\sigma(t)$ underwent a sharp, brief increase at minute \PanicPeakTime, while $\Sigma(t)$ stayed constant within uncertainty.

Because $\Sigma(t)$ showed no systematic variation, in later simulations it was fixed to the baseline value $\Sigma=\Ssigma\Units{cm\,s^{-3/2}}$ already reported in Table~\ref{tab:model-params}. The remaining parameters were summarized by back-to-back exponential profiles centered at $t_c=\PanicPeakTime\Units{min}$,
\begin{equation}
	q(t) = q_0 + A_q \, \dexp{-\lambda_q \lvert t - t_c \rvert},
	\qquad q \in \{\Delta\Sigma, v_\sigma\}
	\label{eq:panic-temporal-fit}
\end{equation}
\noindent
where $q_0$ is the value away from the peak, $A_q$ is the peak excess, and $\lambda_q$ is the common growth/relaxation rate. The six scalar coefficients $(q_0,A_q,\lambda_q)$ for $q=\Delta\Sigma$ and $q=v_\sigma$ are listed in Table~\ref{tab:model-params} and were used in the later simulations. The peak time $t_c$ will subsequently be used as a key measure of one of the most significant timescales in the ants' mechanism in response to panic.

As previously noted, the CM begins to shift almost immediately after repellent introduction (Fig.~\ref{fig:exp-setup}d), with fluctuations masking the earliest rise but a clear upward trend by minute 6, suggesting a response of about 1 minute. A biologically grounded mechanism for this rapid response is recruitment-based information transfer, as observed in tandem communication in ants~\cite{mglich1974tandem}, and it may proceed as follows. Initially dispersed, ants near the barrier likely receive danger cues quickly upon panic induction at minute 5, and become ``activated''. Each then inform to another nestmate after a characteristic time $\tau_f$, doubling the number of activated ants in a cascade that ends at $T = 1 \Units{min}$ with $N_T = 20$ informed individuals. This implies $\dot{N} = N / \tau_f$, with $N(0) = 1$, yielding $\tau_f = T / \ln N_T = 20 \Units{s}$. While multiple ants may simultaneously access the barrier, leading to overlapping information flows, we simplify this dynamics as being driven by a single ``effective ant'' that triggers the cascade, i.e., a representation of the collective's fast, memory-limited reaction to panic cues.

When all ants are informed at minute 6, a slower dynamics arises. It can be characterized by the time between the moment when all members were activated and the peak observed in Fig.~\ref{fig:panic-regime}a at minute \PanicPeakTime, i.e., by a characteristic time $\tau_s = 3.2 \Units{min}$. Hence, an extended SSModel based on Eq.~\eqref{eq:ss-model} that incorporates boundary effects and panic dynamics mediated by the times $\tau_f$ and $\tau_s$, was evaluated as follows:
\begin{subequations}
\begin{align}
	\begin{split}
			\frac{d}{dt} v_i(t) &= 
			-\gamma (v_i(t) - \tilde v_i(t)) \\
			& \quad\, + \sigma(v,t) \sqrt{\hat D(t)} \,\xi_i(t)
			+ F_w(r_i)
	\end{split}
	\label{eq:ss-model-boundary}
	\\
	\sigma(v,t) &= \Sigma(t) + \Delta\Sigma(t) \, (1 - \dexp{-v / v_\sigma(t)})	
	\\
	F_w(r_i) &= a_0 (
		\dexp{-r_i / \ell_w} - 
		\dexp{-(r_i - L) / \ell_w})
	\label{eq:wall-forces}
	\\
	\tilde{\mathbf{v}}(t) &= 
	\left(
		\frac{F_w(\tilde x)}{\gamma}
		\frac{
		1 - \dexp{-(t - t_p) / \tau_s}}{
		1 - \dexp{-(t - t_p) / \tau_f}}, \;0
	\right)
	\label{eq:v-drift}
\end{align}
\label{eq:panic-langevin-model}
\end{subequations}

\noindent
Here, $r_i$ represents the $x$-$y$ position of an ant, and $F_w(r_i)$ in Eq.~\eqref{eq:wall-forces} mimics wall forces with intensity $a_0 = \WallIntensity \Units{cm/s^2}$, as fitted in Fig.~\ref{fig:boundary-effect}a, and arena size $L = \CellLength \Units{cm}$. The drift velocity $\tilde{\mathbf{v}}(t)$ in Eq.~\eqref{eq:v-drift} has zero $y$ component due to the directionality of the panic effect, while its $x$ component reflects fast and slow dynamics with time scales $\tau_f$ and $\tau_s$. It is modeled as an asymmetric sigmoid centered at $t_p = 5.5 \Units{min}$ and saturates at $F_w(\tilde{x}) / \gamma$, balancing wall repulsion and panic drive. The only free parameter, $\tilde{x}$, corresponds to the final cluster position. Fig.~\ref{fig:exp-setup}d shows this extended SSModel reproducing CM motion using experimentally derived parameters, with all values listed in Table~\ref{tab:model-params}.

The diffusion coefficient not only fluctuates on similar timescales before and after panic, but also retains its Log-normal statistics, suggesting a fundamental, stress-resistant feature of confined ant interactions, as reflected in stable acceleration PDFs (Fig.~\ref{fig:after-panic-results}d). Likewise, the near-constancy of $\Sigma(t)$ indicates that danger does not alter the baseline noise intensity, while the return of $\Delta\Sigma(t)$ to its initial value implies that total noise intensity $\Sigma + \Delta\Sigma$ is effectively ``remembered''. This superstatistical framework can be narrated in three stages of response to panic. First, from minute 5 to 6, a \textit{cascade} begins as ants near the barrier become informed and transmit the danger signal across the safe cell, with a characteristic time $\tau_f = \CMTimeSlow \Units{s}$ comparable to the ant memory timescale reported in the literature~\cite{gordon2010ant}. In the second stage, from minute 6 to \PanicPeakTime, the \textit{drift} phase emerges as the CM shifts rightward as informed ants increasingly bias their movement away from the barrier, with a timescale $\tau_s = \CMTimeFast \Units{min}$. Finally, from minute \PanicPeakTime\ to 15, the \textit{cluster} stabilizes as the drift velocity $\tilde{\mathbf{v}}(t)$ saturates, reaching a force balance between the rightward panic repulsion and the leftward mechanical repulsion from the opposite wall.
 \section{Discussion} 

The slowly varying diffusivity $D(t)$ does not necessarily correspond to a single ants physiological knob, instead, it is more naturally interpreted as an emergent combination of several factors that jointly regulate ant mobility under confinement. In this context, $D(t)$ can be viewed as the product of several weakly dependent, positive stochastic factors. Taking the logarithm transforms this multiplicative structure into a sum of random contributions, so that $\log D$ becomes the sum of many small terms. Under these conditions, the Central Limit Theorem applies to  $\log D$, leading to Gaussian statistics for $\log D$ and, consequently, to Log-normal fluctuations of $D$ itself.

At the biological level, these multiplicative stochastic factors can be naturally associated with the mechanisms by which ants communicate and interact while moving under confinement. Ant-ant communication in \textit{Atta insularis} likely involves different ways of short-range chemical communication using various parts of the insect's body \cite{jackson2006communication}, and perhaps also substrate vibrations \cite{hager2017directional}. These would induce a multiplicative contribution to the fluctuating diffusion coefficient: vibrational signals propagate through contact chains, where one information-emitting ant can trigger successive activations in its neighbors, effectively amplifying the signal. Under repeated transmissions, this mechanism is structurally similar to autocatalytic growth, in which successive positive factors are multiplied together \cite{kavanagh1934autocatalytic}. 

Complementarily, from a coarse-grained kinematics perspective, diffusivity can be understood in terms of effective step lengths and times ($D \sim {\ell_{\text{eff}}^2}/{\tau_{\text{eff}}}$), as in generic run-reorient-pause dynamics, where both quantities are themselves composite \cite{tilles2016random,bartumeus2008fractal}. Locomotor vigor, intermittency due to pauses, contact rate, or turning dynamics could gate one another multiplicatively, so mobility can be expressed as a product of positive, slowly fluctuating contributions rather than as an additive control parameter \cite{bartumeus2008fractal,nolting2015composite,feinerman2017individual}.

In confinement, effective space available could be a strong multiplicative constraint, simultaneously modulating encounter rates, turning frequency, and the propensity to pause \cite{gravish2013climbing,berthelot2021random,miller2011movement,estevez2006analysis}. Small fluctuations in local density can therefore trigger a cascade-like process: changes in density alter collision rates \cite{davidson2017spatial}, which in turn modify turning and stop-go statistics, feeding back onto local density itself. Such feedback naturally constitutes a multiplicative cascade in space and time, for which Log-normal statistics are a well known outcome \cite{beck2003superstatistics,mandelbrot1974intermittent,frisch1995turbulence}. In parallel, ant locomotion can be understood as a stochastic switching among internal movement modes (e.g., resting, slow cruising, fast cruising), with both the residence time in each mode and the transition rates between modes influenced by local encounters \cite{richardson2021ant,gallotti2018ants}. The resulting effective mobility then emerges as a product of locomotor state, time spent in that state, and interaction-mediated gating, again pushing toward Log-normal $D$.

Within this picture, the emergence of Log-normal diffusivity is not driven by transitions between distinct behavioral programs at a colony level (e.g., baseline exploration versus panic response), but by interaction-mediated processes that continuously reshape individual trajectories at the collective level. Panic, therefore, does not create a new dynamical regime; instead, it transiently perturbs a system that is already governed by multiplicative processes. Because the possible multiplicative sources discussed above remain active during panic, the mechanisms responsible for generating Log-normal fluctuations are preserved, endowing the system with a form of statistical resilience. In this sense, confinement could act less as a passive boundary condition and more as a regulator of collective stochasticity.

A remaining issue is the contrast between the long-term stability of the cluster despite ants' short individual memory. One possible explanation is that the cluster is not entirely static: some ants periodically leave the cluster, explore the barrier, and return (see video at SI). We hypothesize that the information feedback provided by these individuals is rapidly transmitted to the rest of the densely-packed ants, so the cluster-formation behavior is constantly reinforced. This idea is supported by Fig.~\ref{fig:panic-regime}b, which shows a drop but not a disappearance in barrier exploration probability following repellent injection. As the cluster stabilizes, its left edge remains roughly $2 \Units{cm}$ from the barrier. The sustained but reduced traffic suggests that returning ants subtly refresh the danger signal, helping maintain cluster cohesion during the final 5 minutes of the experiment. \section{Conclusions}

In summary, we have shown that Log-normal Superstatistics emerges as a robust statistical framework that captures the transition from normal to danger behavior in confined ants, while maintaining continuity and providing a foundation upon which panic-induced dynamics can be superimposed. We reported non-Gaussian behavior, a slowly varying diffusion coefficient, and a collective escape response consistent with interaction-mediated information transfer. A data-driven superstatistical model with Log-normal diffusion successfully reproduced these features and captured the danger transition, relying on a single free parameter.

The model uncovers a fast, memory-limited information cascade followed by a slower collective reorganization into a stable cluster, and strongly suggests that the diffusivity itself is not a single physiological control parameter but an emergent quantity arising from the multiplicative combination of interaction-mediated processes under confinement. When expressed as a product of weakly dependent stochastic factors, the logarithm of the diffusivity becomes additive, naturally yielding Log-normal fluctuations via the Central Limit Theorem. The persistence of this statistical structure under panic indicates that the underlying multiplicative mechanisms remain intact, revealing a form of collective statistical resilience and positioning confinement as a regulator of stochastic organization in living active matter.
 \section{Acknowledgements} 

A. Reyes gratefully acknowledges the Abdus Salam ICTP STEP program for financial support. We thank A. Celani, D. Gordon, R. Mulet, T. Tohme and D. Machado for valuable discussions, G. Simon for the 3D-printed of the cells, and A. Serrano for assistance with the design of Fig.~\ref{fig:exp-setup}. E. Altshuler acknowledges an Invited Professorship supported by CNRS at the University of Lyon.
 \begin{figure}
	\includegraphics{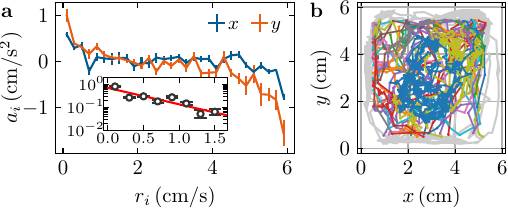}
	\caption{
		Boundary Effects and Effective Wall Thickness.
		(a) Cartesian acceleration components as functions of Cartesian position, with error bars showing the standard error of the mean. Inset: acceleration absolute value as a function of the distance from the walls. A characteristic length $\ell_w=\WallExtent\,\mathrm{cm}$ is defined as the effective wall thickness.
		(b) The grey sections have been removed from a sample trajectory, as they are within a distance $\ell_w$ from the walls: in order to ensure that statistics is not contaminated by boundary effects, we only use for analysis the non-gray sections.}
	\label{fig:boundary-effect}
\end{figure}

\section*{Appendix A: Boundary effects}

The baseline regime was used to evaluate boundary effects while measuring the extent of wall ``forces''. Experimental Cartesian acceleration components (i.e., $a_x$ and $a_y$) were arranged based on increasing values of the corresponding Cartesian position components $x$ and $y$, respectively. Both acceleration time series were binned and a running average was taken (main panel of Fig. \ref{fig:boundary-effect}a). The ants slow down as they approach the walls, revealing the influence of the boundaries as if they were confined by a soft, square-box-like potential. Acceleration absolute value against the distance from the wall revealed a characteristic length $\ell_w=\WallExtent\Units{cm}$, obtained by fitting an exponential function (inset of Fig. \ref{fig:boundary-effect}a). Henceforth, analysis was conducted with trajectory segments that were consistently at least $\ell_w$ units away from every boundary, to discard boundary effects (Fig. \ref{fig:boundary-effect}b).

\begin{figure}
	\includegraphics[width=\linewidth]{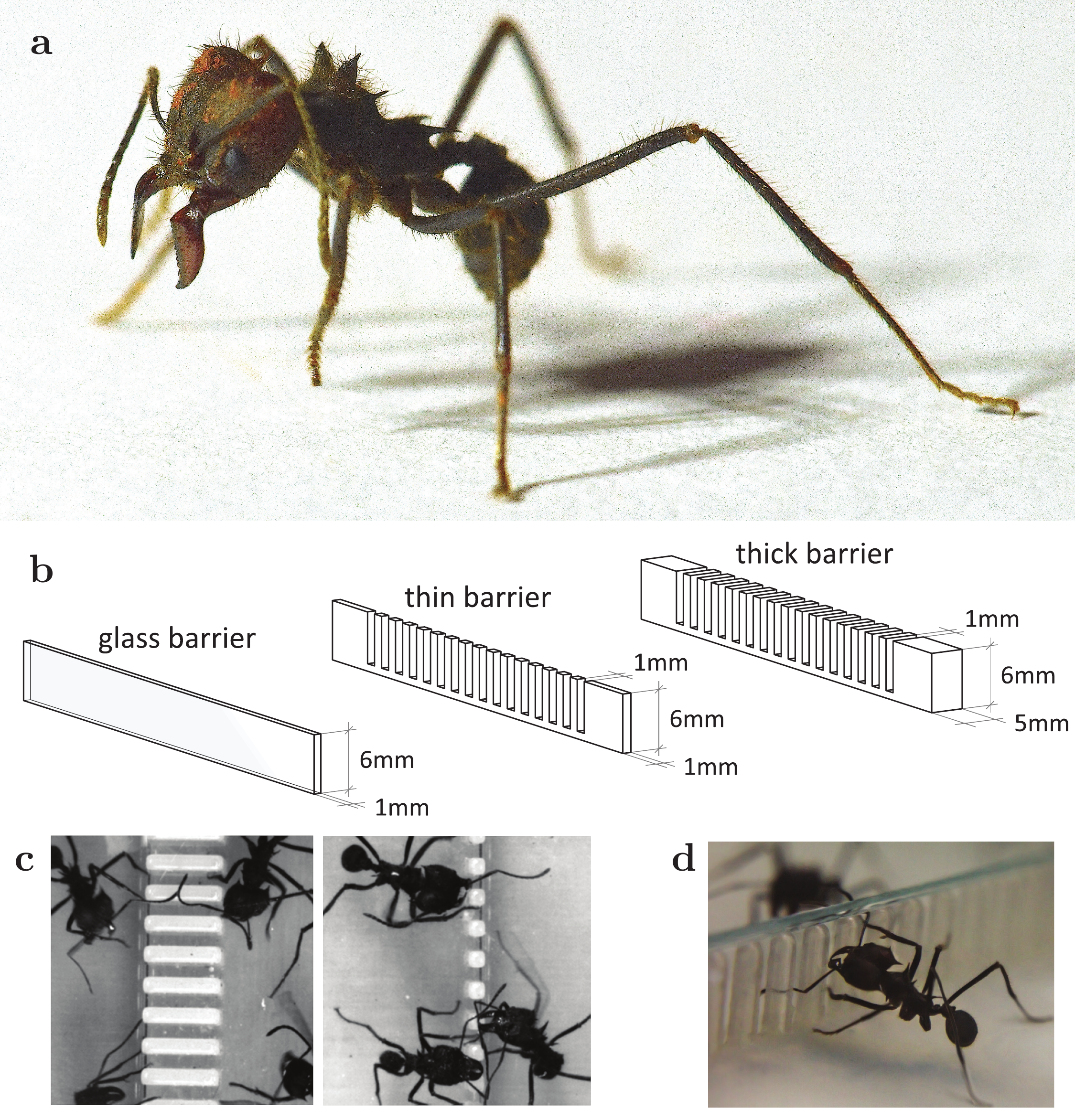}
    \caption{
		Barrier Designs and Contact Pathways for Danger Transmission.
		(a) Ant used in experiments: leafcutter ant \textit{Atta insularis}, a species native to Cuba known as ``bibijagua''.
		(b) Sketch of the three types of barriers used for the protocols \textit{Glass}, \textit{Thick} and \textit{Thin}. The latter is also used for the \textit{Ctrl} protocol. 
		(c) Top view of ants near a thick and thin barriers. Short distance contacts are not allowed by the thick barrier. 
		(d) Lateral view of an ant near a thin barrier. As a reference, the length of the ants bodies is around $1\Units{cm}$.}
	\label{fig:atta-insularis}
\end{figure}

\section*{Appendix B: Mechanisms for danger transmission}

To isolate potential mechanisms of danger transmission, we used three barrier types (Fig.~\ref{fig:atta-insularis}b): a transparent \textit{Glass} wall allowing only visual cues; a \textit{Thick} barrier restricting the interaction to mutual antennation; and a \textit{Thin} barrier allowing full contact between ants in adjacent arenas (Fig.~\ref{fig:atta-insularis}c).

\begin{figure*}[t]
	\noindent
	\begin{minipage}[t]{\dimexpr0.5\textwidth-0.5\columnsep\relax}
		\vspace{0pt}
		\centering
		\includegraphics[width=\linewidth]{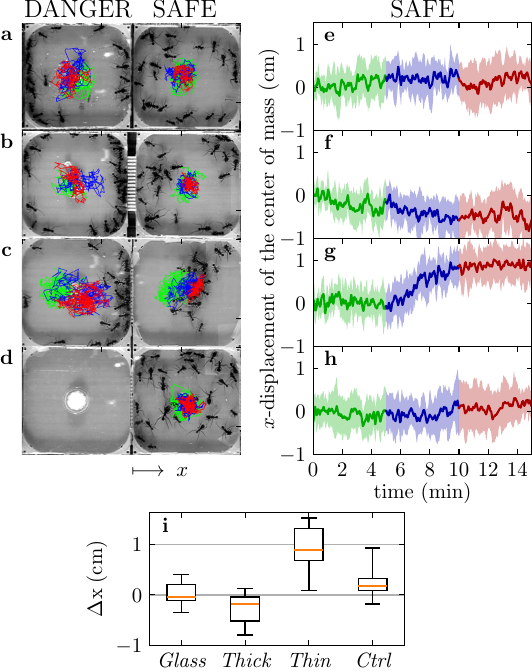}
	\end{minipage}\hfill
	\begin{minipage}[t]{\dimexpr0.5\textwidth-0.5\columnsep\relax}
		\vspace{0pt}
		\caption{
			Barrier-Dependent Panic Transmission and Center of Mass Analysis.
			Snapshots of experiments 
			(a) \textit{Glass}, 
			(b) \textit{Thick}, 
			(c) \textit{Thin} and 
			(d) \textit{Ctrl}. Green, blue, and red traces show the center of mass (CM) trajectory for consecutive 5-minute windows. 
			(e-h) Horizontal CM evolution in the safe cell, relative to its initial position. Shading shows standard deviation. Repellent is added to the danger cell at $t=5$ min.
			(i) Net horizontal displacement of the ants CM based on 10 repetitions of each type of experiment. Mann-Whitney tests assert to the \textit{Thin} protocol to statistically differs from the other three.
		}
		\label{fig:cm-displacements}
	\end{minipage}
\end{figure*}

To characterize collective motion across protocols, we tracked the ants' CM in both arenas (Fig.~\ref{fig:cm-displacements}a-d). In all cases, the CM in the danger cell remained near the geometric center regardless of barrier type. The same held for the safe cell, except under the \textit{Thin} protocol, where the CM systematically shifted away from the barrier after repellent was introduced (Fig.~\ref{fig:cm-displacements}c). Fig.~\ref{fig:cm-displacements}e-h shows the time evolution of the horizontal CM position in the safe cell, where mean and standard deviation are shown as a function of time, averaged over 10 experimental repetitions. A clear displacement emerges only in the \textit{Thin} case after the injection at minute 5. In contrast, the \textit{Ctrl} case (i.e., where the danger cell is empty) shows no comparable motion, indicating that passive diffusion alone cannot account for the observed shift.

To quantify this, we computed the net CM displacement in the safe cell as the average difference in $x$-coordinate between minutes 5 and 15 of each experiment (Fig.~\ref{fig:cm-displacements}i). We performed one-tailed Mann-Whitney U tests to evaluate the null hypothesis that each pair of samples (\textit{Thin} vs. \textit{Glass}, \textit{Thick}, and \textit{Ctrl}) originates from equal populations and used a significance level of $\alpha = 0.05$. We obtained $U = 84.0$, $89.0$, and $71.0$, and corresponding $p$-values of $0.0008$, $0.0002$, and $0.0187$, respectively. Since all $p$-values fall below the threshold, the null hypothesis is rejected in each case.

These results indicate that only the \textit{Thin} barrier enables the transmission of panic-inducing information sufficient to trigger a significant escape response in ants within the safe cell. The \textit{Ctrl} data reinforce this, showing that repellent diffusion alone, without ant-to-ant signaling, has no effect. Minimal shifts in the \textit{Glass} and \textit{Thick} protocols further suggest that neither visual cues nor simple antennation are effective mechanisms for danger transmission, but touching other body parts seems necessary.

\begin{table}[b]
	\centering
	\caption{Values for model parameters fitted to data.}
    \label{tab:model-params}
	\begin{tabular}{c c c c c c}
		\hline\hline
		$\ell_w$ & $a_0$ & $\gamma^{-1}$ & $\Sigma$ & $\Delta \Sigma$ & $v_\sigma$
		\\
		$\mathrm{(cm)}$ & $\mathrm{(cm/s^2)}$ & $\mathrm{(s)}$ & $\mathrm{(cm\,s^{-3/2})}$ & $\mathrm{(cm\,s^{-3/2})}$ & $\mathrm{(cm/s)}$
		\\
		\WallExtent & \WallIntensity & \InvGamma & \Ssigma & \DeltaSigma & \Vc
		\\
		\hline
		$s$ & $\tau_f$ & $\tau_s$ & $L$ & $t_p$ & $\tilde{x}$
		\\
		-- & $\mathrm{(s)}$ & $\mathrm{(min)}$ & $\mathrm{(cm)}$ & $\mathrm{(min)}$ & $\mathrm{(cm)}$
		\\
		\ShapeParam & \CMTimeSlow & \CMTimeFast & \CellLength & \TStartSigmoid & \CMSaturationLength
		\\
		\hline
		$\Delta\Sigma_0$ & $A_{\Delta\Sigma}$ & $\lambda_{\Delta\Sigma}$ & $v_{\sigma_0}$ & $A_{v_\sigma}$ & $\lambda_{v_\sigma}$
		\\
		$\mathrm{(cm\,s^{-3/2})}$ & $\mathrm{(cm\,s^{-3/2})}$ & $\mathrm{(s^{-1})}$ & $\mathrm{(cm/s)}$ & $\mathrm{(cm/s)}$ & $\mathrm{(s^{-1})}$
		\\
		\PanicDeltaSigmaBase & \PanicDeltaSigmaAmp & \PanicDeltaSigmaRate & \PanicVcBase & \PanicVcAmp & \PanicVcRate
		\\
		\hline\hline
	\end{tabular}
\end{table}

\onecolumngrid

\vspace*{\fill}
\vspace*{\fill}

\clearpage
\twocolumngrid

\end{document}